\documentclass[aps,prb,twocolumn,showpacs,floatfix]{revtex4-1}

\usepackage[caption=false]{subfig}
\usepackage{verbatim}
\usepackage{graphicx}
\usepackage{dcolumn}
\usepackage{bm}
\usepackage{color}
\usepackage[colorlinks=true, allcolors=blue]{hyperref}
\usepackage{makeidx}
\usepackage{amsmath}
\usepackage{amssymb}
\makeindex
\def\>{\right\rangle}
\def\<{\left\langle}
\def\be{\begin{equation}}
\def\ee{\end{equation}}
\def\ba{\begin{array}{l}}
\def\ea{\end{array}}

\def\beq{\begin{eqnarray}}
\def\eeq{\end{eqnarray}}

\newcommand{\bra}[1]{\langle #1 |}
\newcommand{\ket}[1]{| #1 \rangle}
\def\trace{\text{Tr}}
\newcommand{\meanOmega}[1]{\left\langle #1 \right\rangle_{\!\Omega}}

\begin{document}
\title{Time-{resolved} energy dynamics after single electron injection into an interacting helical liquid}
\author{Alessio Calzona$^{1,2}$, Matteo Acciai$^1$, Matteo Carrega$^2$, Fabio Cavaliere$^{1,2}$, and Maura Sassetti$^{1,2}$}
\affiliation{ $^1$ Dipartimento di Fisica, Universit\`a di Genova,Via Dodecaneso 33, 16146, Genova, Italy.\\
$^2$ SPIN-CNR, Via Dodecaneso 33, 16146, Genova, Italy.
} 
\date{\today}
\begin{abstract}
The possibility to inject a single electron into ballistic conductors is at the basis of the new field of electron quantum optics. Here, we consider a single electron injection into the helical edge channels of a topological insulator. Their counterpropagating nature and the unavoidable presence of electron-electron interactions dramatically affect the time evolution of the single wavepacket. Modeling the injection process from a mesoscopic capacitor in presence of non-local tunneling, we focus on the time resolved charge and energy packet dynamics. Both quantities split up into counterpropagating contributions whose profiles are strongly affected by the interactions strength. In addition, stronger signatures are found for the injected energy, which is also affected by the finite width of the tunneling region, {in contrast to what happens for the charge}. Indeed, the energy flow can be controlled by tuning the injection parameters and we demonstrate that, in presence of non-local tunneling, it is possible to achieve situation in which charge and energy flow in opposite directions. 
\end{abstract}
\maketitle
\section{Introduction}
Electron-electron (e-e) interactions in one dimensional systems play a prominent role \cite{giamarchi2003quantum, vondelft}. The celebrated Fermi liquid theory dramatically fails and intriguing phenomena appear, such as the fractionalization of the charge \cite{deshpande2010electron, barak2010interacting, maslov1995landauer, safi1995transport,pham2000fractional} and the spin  \cite{pham2000fractional,calzona2015spin, das2011spin, garate2012noninvasive, calzona2015physe} degrees of freedom. Here, an electron injected into an interacting system splits up originating two collective excitations which carry a fraction of the electron charge and spin. In this context, many other theoretical predictions have been put forward \cite{maslov1995landauer, safi1995transport, safi1997properties, calzona2015spin} and, recently, some of them have also been experimentally tested \cite{bena2001measuring, steinberg2007charge, kamata2014fractionalized}. Among all, it is worth to mention the direct observation of charge fractionalization in chiral conductors by means of time-resolved charge current measurements reported by Kamata et al., see Ref.\onlinecite{kamata2014fractionalized, perfetto2014time}.

Despite the great interest on fractionalization phenomena, up to now little attention has been devoted to the study of the energy associated to electrons injected into an interacting system. In Ref.~\onlinecite{karzig2011energypart} it has been shown that the DC energy current along a quantum wire is partitioned between left- and right- moving excitations, but in a distinct way with respect to that of the injected charge. In particular, it has been shown that, differently from the 
charge, the energy partitioning depends on the injection process and its evidence can be already tested in a DC configuration. 

Nevertheless, an accurate description and understanding of energy dynamics for time dependent single electron injection
in an interacting system is still lacking, despite it will play an important role  for the fast developing field of {\em electron quantum optics} \cite{grenier2011electronoptics,bocquillon2014electron}. This very promising field relies on the possibility of injecting single electrons and holes into one dimensional (1D) systems. On-demand single electron sources can be experimentally realized by means of driven mesoscopic capacitors  \cite{feve2007mesoscopic,mahe2010mesoscopic,buttiker1993mesoscopic,moskalet2008mesoscopic} or properly designed Lorentzian voltage pulses  \cite{dubois2013lorentzian,grenier2013lorentzian,dubois2013lorentzianNature, ferraro14prl}.

Injected wave packets propagate ballistically along 1D systems such as integer quantum Hall edge states,  allowing for optics-like experiments where e.g. quantum point contacts act as the analog of beam splitters. In this regard it is worth mentioning two seminal experiments, based on the chiral edge state of a $\nu=2$ quantum Hall system, dealing with the so-called Hanbury-Brown-Twiss  \cite{bocquillon2012electron} and the Hong-Ou-Mandel  \cite{bocquillon2013hom} effects. Different theoretical works have investigated single electron injection in chiral conductors~\cite{grenier2011electronoptics, buttiker1993mesoscopic, moskalet2008mesoscopic} and the role of e-e interactions in copropagating edge channels~\cite{sukhorukov2016prb, ivan2012prb, wahl2014prl, rech12, ferraro15}, aiming to the explanation of recent experimental observations. In addition, the heat and energy transport has been also considered\cite{moskalets1, moskalets2, moskalets3} in presence of external drive but {\em only} in absence of e-e interactions.

 Recently there have been suggestions that counterpropagating helical edge states of two-dimensional topological insulators (2DTI)  \cite{hasan2010colloquium,qi2011topological} can also be used as electronic wave guides. They can be realized in CdTe/HgTe  \cite{bhz2006,dolcettoreview, konig2007quantum} and InAs/GaSb  \cite{liu2008inasgasb,lingjie2015inasgasb,knez2011inasgasb} quantum wells. Importantly, they are topologically protected from backscattering and characterized by the so called spin-momentum locking. These features allow for a richer phenomenology, in comparison with quantum Hall-based setups  \cite{Li15, ferraro14}, and for the study of effects related to spin-entanglement \cite{inhofer2013,hofer2013,strom2015entaglemntspin}, relevant for quantum computation implementations.
In this context e-e interactions between counterpropagating edge channels can lead to remarkable effects, also in comparison to the case of interacting copropagating edge states of chiral conductors \cite{sukhorukov2016prb, ivan2012prb, wahl2014prl, rech12, ferraro15}. A deep understanding of the role of interactions is thus of great importance in view of all these realizations.

In this work we consider the on-demand injection of a single polarized electron from a quantum dot (QD) mesoscopic capacitor into a couple of interacting helical edge states, modeled as helical Luttinger liquid (HLL)  \cite{luttinger1963exactly,tomonaga1950remarks, Wu06, zheng11, Li15}.
Our goal is to study how the presence of e-e interactions affects the dynamics after an injection of a single electron into the edge channels of a 2DTI. In particular, we will focus on the time evolution of both charge and energy densities in the ballistic helical conductor. We will demonstrate that both quantities display fractionalization phenomena, due to the  presence of e-e interactions, resulting in left- and right- moving charge and energy profiles. Interestingly, the time-evolution of the injected energy presents different features that strongly depend on the nature of the tunneling process, in sharp contrast to what happens for the charge degree of freedom.
We will investigate these features, considering that the quantum dot has finite dimension and  thus allowing for non-local tunneling process.
In this case it is also possible to achieve situations in which the charge and energy packets flow in opposite directions, simply by tuning external parameters (such as gate voltages).
Our work will shed new lights on interaction effects in 1D systems, extending previous results obtained only in the DC regime and in the asymptotic limits of local and very extended tunneling have been investigated in Ref.~\onlinecite{karzig2011energypart}..

The manuscript is organized as follows.
In Sec. \ref{sect:model} we describe the setup, presenting  the time dependent density matrix approach to single electron tunneling injection. The charge density is investigated in Sec.~\ref{sect:charge}, where charge fractionalization factors and  time evolution of the single wave packet are derived. Here, the case of local injection is discussed in detail.
Sec.~\ref{sec:energy} is devoted to the study of the energy dynamics. Here, we focus on energy density profiles and on energy partitioning, highlighting the role of e-e interactions and finite width of the tunneling region. 
 
\section{Setup and general model}
\label{sect:model}

\begin{figure}[t]
	\includegraphics[width=\columnwidth]{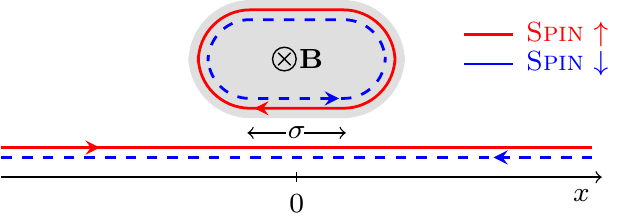}
	\caption{(color online) Sketch of the setup. A mesoscopic capacitor, quantum dot, is tunnel coupled to the helical edge states of a 2DTI trough an extended tunneling region of width $\sigma$. Solid red lines refer to 1D non-interacting electron states with spin-up. Dashed blue lines are associated to spin-down electrons. By means of a top gate (in light gray) it is possible to shift the quantum dot energy levels. Their spin degeneracy can be broken with a magnetic field $\mathbf{B}$ present in the QD region and perpendicular to its plane.}
	\label{fig1}
\end{figure}
We consider helical edge channels (EC) of a 2DTI tunnel coupled with a quantum dot acting as a mesoscopic capacitor, as schematically shown in Fig.~\ref{fig1}. The QD can be realized by means of metallic gates separating the island region from the 2DTI or by means of mechanical etching, in close analogy with what was done in quantum Hall based devices  \cite{feve2007mesoscopic}.
The presence of electrostatic gates screens e-e interactions in the QD region \cite{hofer2013, inhofer2013} with energy levels dominated by confinement rather than Coulomb charging energy. {Infact, it has been experimentally shown that the presence of a top gate in a mesoscopic capacitor results in a strong suppression of e-e interactions, with a very small charging energy contribution~\cite{bocquillon2014electron, feve2007mesoscopic, fevenatcomm, fevephyse}. This explains the success \cite{feve2007mesoscopic} of non-interacting models usually considered in describing the QD region. Motivated by these experimental findings we will consider the QD as non-interacting and we will also neglect possible interaction contributions between the dot and the helical edge states, in view of the presence of electrostatic gates.\cite{note_interactions}}. Due to its finite dimension $l$, the QD has discrete single particle energy levels that come in Kramers pairs and are spaced by $\Delta \sim v_{\rm F}/l$, where $v_{{\rm F}}$ is the Fermi velocity  (throughout this paper we set $\hbar=1$).

The energy spectrum of the QD can be tuned and shifted (with respect to the Fermi energy $E_{{\rm F}}$ of the whole system) by properly acting on a top gate. Moreover, spin degeneracy can be lifted  \cite{hofer2013} by means of a perpendicular magnetic field $\mathbf{B}$ in the island region (see Fig. \ref{fig1}). The injection of a single polarized electron into the EC can be thus achieved with an abrupt change of the top gate potential at time $t=0$. Consequently, the most energetic electron in the QD, chosen here with spin up, is suddenly brought above $E_{{\rm F}}$ and leaves it  \cite{hofer2013,bocquillon2014electron} tunneling into the helical edges. {In this paper, we will assume that the spin-preserving tunneling is the dominant mechanism and we thus restrict the discussion to this case.}

The Hamiltonian of the whole system reads
\begin{equation}
\hat H = \hat H_{EC} +\hat H_{QD} + \hat H_t
\end{equation}
where the edge channels term is $\hat H_{EC} = \hat H_0 + \hat H_{e-e}$. Here the free Hamiltonian reads\begin{equation}
\hat H_0 = v_{{\rm F}} \! \int \! dx \left[\hat \psi^\dagger_L (x)\, i\partial_x \,\hat \psi_L(x) -  \hat \psi^\dagger_R (x) \,i\partial_x \,\hat \psi_R(x)  \right]
\end{equation}
where $\hat \psi_r(x)$ is the fermionic field annihilating electrons in the right- ($r=R$) or left- ($r=L$) branches. As shown in Fig. \ref{fig1}, we consider R-electrons (L-electrons) having spin up (down). The presence of short-range e-e interactions can be taken into account by the additional contribution
\begin{equation}
\hat{H}_{e-e} = \frac{g_4}{2} \sum_r \! \int \! dx \left(\hat n_r(x)\right)^2 + g_2 \int\! dx \; \hat n_R(x) \hat n_L(x),
\end{equation}
with
\begin{equation}
\label{eq:particle_fermionic}
\hat n_r(x) = :\hat \psi_r^\dagger(x) \hat \psi_r(x): 
\end{equation}
the electron density on the $r$ channel and $g_i$ coupling constants referring to inter- ($i=4$) and intra-channel ($i=2$) interactions \cite{vondelft, miranda2003, sassetti96}. 

The QD is represented in terms of a spin-up electron level  $\epsilon_0$, measured with respect to the Fermi energy,
\begin{equation}
\hat{H}_{QD} = \epsilon_0 \hat d^\dagger \hat d\,.
\end{equation}
This situation can be achieved by a sudden shift of the uppermost occupied electron above $E_{{\rm F}}$, with $\epsilon_0>0$  bounded by the level spacing $\Delta$ \cite{hofer2013}.

The tunneling of the spin-up electron between the dot and the helical EC is represented by $\hat H_t = \hat H_t^+ + \hat H_t^-$ where
\begin{equation}
	\label{eq:tunneling}
	\hat H_t^+ = \lambda \int_{-\infty}^{+\infty}\!\!dy\;w(y) \, \hat \psi_R^\dagger(y) \hat d \,,\quad 	\hat H_t^- =  (\hat H_t^+)^\dagger
\end{equation}
respectively adds or removes a spin-up electron to the edge right channel with constant tunneling amplitude $\lambda$. 
The tunneling region is characterized by the envelope function $w(y)$, whose precise shape will be specified later\cite{chirolli2011extended, chevalier12, dolcetto12}. We assume $w(y)$ centered around $y=0$ (the injection point) with a spatial extension given by $\sigma$. 
For geometrical reasons (see Fig. \ref{fig1}), $\sigma$ is bounded by the QD dimension  $\sigma<l$ and thus, in view of the constraint $\epsilon_0<\Delta$, also by $\sigma\lesssim v_{{\rm F}}/\epsilon_0$. 

Concerning the momentum involved in the tunneling process, the bottom edge of the QD (described as a system of spin up and left moving electrons) has momentum $k_{QD}$, while the right electrons in the helical EC have momentum $k_{{\rm F}}$. One can then define the total variation as $k_0=k_{QD}-k_{{\rm F}}$. This quantity can be tuned by means of gate voltages applied to the QD and/or to the edge channels~\cite{dolcini2011prb, citro2011prb, romeo2012prb}. It will be incorporated into the envelope function $w(y)$ as a complex phase factor $w(y)= {\xi}(y)e^{i k_0 y}$, with $\xi(y)$ real. This phase factor plays a relevant role in the case of non-local tunneling, as we will show. Note that in the case of local tunneling, with $\sigma\to 0$, one has  $w(y)=\xi(y)=\delta(y)$.

\subsection{Single electron injection}
In this section we model the single electron injection process. Assume that at time $t=0$ the edge channels of the 2DTI are at thermal equilibrium (at temperature $T$) with fixed particle number $N$ and with an equilibrium density matrix $\hat\rho_{EC}(0)$. 
On the contrary, the single QD level is initially occupied and described by the density matrix $\hat\rho_{QD}(0) = \ket{1}\bra{1}$.

Let $\hat O(x)$ be a generic hermitian and number-conserving operator that acts on the EC,  such as, for example, the particle density or the Hamiltonian density of the edge channels. In the interaction picture, with respect to the tunneling Hamiltonian $\hat{H}_t$, the time evolution average of $\hat O(x,t)$ reads
\begin{equation}
\langle \hat O(x,t)\rangle = \trace\left\{\hat O(x,t) \hat\rho(t)\right\}\,,
\end{equation}
with the time dependent density matrix
\begin{equation}
\hat\rho(t) =\hat  U(t,0) \hat\rho(0) \hat U^\dagger(t,0) 
\end{equation}
where
\begin{align}
\hat U(t,0) &= \mathcal{T} \left[e^{-i \int_0^t dt' \hat H_t(t')}\right] \\
\hat\rho(0) &= \hat\rho_{EC}(0) \otimes \hat\rho_{QD}(0).
\end{align}
We are interested in the average variation of $\hat{O}(x,t)$ induced by the tunneling process, defined as {$\delta O(x,t)=\langle \hat{O}(x,t)\rangle - \trace\left\{ \hat O (x,t)\hat\rho(0) \right\}$}.
 At lowest order in the tunneling one has
\begin{equation}
\label{eq:DeltaO_general}
\begin{split}
&\delta O(x,t)  = \\
&= 2\Re \!\int_0^t\!\!dt_2\!\!\int_0^{t_2}\!\!dt_1\; \trace\left\{\hat \rho(0) \hat H_t^-(t_1) \left[\hat O(x,t), \hat H_t^+(t_2)\right]\right\}_{\!1,N}
\end{split}
\end{equation}
where the symbol $\trace\left\{\dots\right\}_{1,N}$ denotes the trace over system's excitations with fixed particle numbers: one electron in the QD and $N$ in the EC.

In the following we will consider the low temperature limit (temperature smaller than the energy level splitting of the QD and of the energy excitations of the helical edge), setting $T\to 0$ and thus $\hat\rho_{EC}(0) = \ket{\Omega_N}\bra{\Omega_N}$, with $\ket{\Omega_N}$ the N-particles EC ground state.
Moreover we explicitly take into account the finite lifetime {$1/2\gamma$} of the QD level \cite{bocquillon2014electron, martin2014} by assuming the time evolution of QD {correlator
$\langle\hat d^\dagger(t_1)\hat d(t_2)\rangle=\beta^*(t_1)\beta(t_2)$}, with 
\begin{equation}
\beta(t) =  e^{-i \epsilon_0 t} e^{-\gamma t}.
\end{equation}
{The parameter $2\gamma$ describes the inverse lifetime of the electron in the QD.} {Its precise value will be microscopically calculated in Sec. \ref{sect:lifetime} exploiting again the time evolution of the density matrix.}
In any case, the \emph{single} electron injection implies that the QD level is sufficiently well-defined, with $\gamma$ smaller than both the level position $\epsilon_0$ and the spin level splitting. All these assumptions allow to express $\delta O(x,t)$ as 
\begin{equation}
\label{eq:averageO}
\delta O(x,t) = |\lambda|^2 \; 2 \Re \int_0^t\!\!dt_2\!\!\int_0^{t_2}\!\!dt_1\!\!\iint_{-\infty}^{+\infty}\!\!dy_1dy_2\;\;  \Xi\;  \mathcal{I}_O
\end{equation}
where 
\begin{align}
\Xi(t_1,t_2,y_1,y_2) &=\beta^*(t_1)\beta(t_2)w^*(y_1)w(y_2)\,, \\
\label{eq:I}
\begin{split}
\mathcal{I}_O(t_1,t_2,y_1,y_2,t,x) &=\mathcal{I}_O^{(a)}+\mathcal{I}_O^{(b)}\\
&\hspace{-4em} = \langle\hat  \psi_R(y_1,t_1) \left[\hat O(x,t),\hat \psi_R^\dagger(y_2,t_2)\right]\rangle_\Omega.
\end{split}
\end{align}
Here $\langle \dots \rangle_\Omega$ is a shorthand notation for the ground state average $\bra{\Omega_N} \dots \ket{\Omega_N}$.

It is interesting to briefly discuss the two terms in Eq. \eqref{eq:I}. Once the injection is ended, i.e. for $t\gg(2\gamma)^{-1}$, the first term 
\begin{equation}
\mathcal{I}_O^{(a)} = \langle \hat \psi_R(y_1,t_1) \hat O(x,t) \hat \psi_R^\dagger(y_2,t_2) \rangle_\Omega
\end{equation}
gives a contribution to $\delta O(x,t)$ that can be always expressed as an average over  a pure quantum state  $\ket{S}$ of $N+1$ electrons, namely 
\begin{equation}
\delta O^{(a)}(x,t) = |\lambda|^2 \bra{S} \hat O(x,t) \ket{S}\,,
\end{equation}
with  
\begin{equation}
\ket{S} = \int_0^{\infty} \!\!dt\!\!\int_{-\infty}^{+\infty} \!\!dy\; \beta(t)w(y) \; \hat \psi_R^\dagger(y,t) \, \ket{\Omega_N}.
\end{equation}
Note that in presence of e-e interactions and counterpropagating modes, the field operator $\hat \psi_R^\dagger(y,t)$ is {\em not} chiral.
{Therefore the state $\ket{S}$ cannot be expressed as single integral over space or time unless the injection is local, with $w(y)=\delta(y)$.}
The other term $\mathcal{I}_O^{(b)}$, instead, cannot be expressed as an average over a pure quantum state. We will see that it does not contribute to the total injected charge and energy, but it induces fluctuations in the charge and energy density profiles at fixed $N$ electrons.

\subsection{Dealing with e-e interactions}
Electron-electron interactions can be properly handled using well-known bosonization techniques  \cite{vondelft, miranda2003}. The interacting helical Hamiltonian can be diagonalized introducing proper chiral bosonic fields $\hat \phi_\eta(x,t)$, with $\eta=\pm$ referring to the direction of propagation  (right and left respectively). Fermionic $r$-fields $\hat \psi_r(x,t)$ can be expressed in terms of bosonic ones $\hat \phi_r(x,t)$ as (omitting Klein factors and considering $\vartheta_{R,L}=\pm 1$)
\be
\label{eq:boson_eq}
\hat \psi_r^\dagger (x,t) = \frac{1}{\sqrt{2 \pi a}}  e^{i \sqrt{2 \pi} \hat \phi_r(x,t)} \; e^{-i \vartheta_r k_{\rm F} x}
\ee
with $a$ the usual short-length cut-off  \cite{vondelft, miranda2003, notecutoff}. The complex phases $e^{\pm i k_{{\rm F}}x}$ present in the above expression will play a role {\em only} in $\hat{H}_t$. As discussed above, we already took into account these contributions introducing a complex phase in the definition of the envelope tunneling function $w(y)= \xi(y) e^{i k_0 y}$.

The boson fields $\phi_r(x,t)$ are related to the chiral ones $\hat{\phi}_\pm(x,t)$ by
\begin{subequations}
	\label{eq:r_to_eta}
	\begin{align}
	\hat{\phi}_R(x,t) &= A_+ \hat{\phi}_+(x,t) + A_- \hat{\phi}_-(x,t) \\
	\hat{\phi}_L(x,t) &= A_- \hat{\phi}_+(x,t) + A_+ \hat{\phi}_-(x,t)
	\end{align}
\end{subequations}
where 
\be
A_\pm= \frac{1}{2} \left(\frac{1}{\sqrt{K}} \pm \sqrt{K}\right)
\ee
contain the HLL interaction parameter $K$ 
\be
K = \sqrt{\frac{2 \pi v_{\rm F} +g_4 -g_2}{2 \pi v_{\rm F} +g_4 +g_2}}.
\ee
 Since we consider the case of very long EC, hereafter we can safely neglect the contribution describing zero modes \cite{vondelft}. 
The Hamiltonian can be written in diagonal form as 
\be
\hat{H}_{EC} = \int_{-\infty}^{\infty} \mathcal{\hat{H}}(x,t) \, dx\,,
\ee
with the Hamiltonian density associated to boson collective modes given by
\be
\label{eq:def_H}
\mathcal{\hat{H}}(x,t) = \frac{u}{2} \sum_{\eta=\pm} :(\partial_x \hat \phi_\eta(x-\eta u t))^2:\,.
\ee
Here $u= (2 \pi)^{-1} \left[(2\pi v_{\rm F} + g_4)^2 - (g_2)^2\right]^{1/2} $ represents the renormalized propagation velocity. For the sake of simplicity, we will consider $u = v_{\rm F} K^{-1}$ that holds as long as $g_2=g_4$. However, other kinds of repulsive interactions, possible in a helical EC, can be straightforwardly taken into account.

\subsection{Inverse lifetime}
\label{sect:lifetime}
{We now evaluate the inverse lifetime $2\gamma$ of the QD level at the lowest order in the tunneling.
Recalling that the system is initially in a state with one electron in the dot and $N$ electrons in the edge channels, the transition probability
is given by the relation $P_{1 \to 0}(t)=\mathrm{Tr}\{\bra{N+1,0}\hat\rho(t)\ket{N+1,0}\}$, where $\ket{N+1,0}$ denotes the state with no electrons in the dot and $N+1$ electrons in the edge
channels. The trace is calculated over the excitations of the system at fixed particle number. At lowest order in the tunneling one has
\begin{equation}
\begin{split}
P_{1\to 0}(t)&=|\lambda|^2\iint_0^t dt_1dt_2\iint_{-\infty}^{+\infty}\!\!dy_1dy_2e^{i\epsilon_0(t_1-t_2)}\\
		&\quad w^*(y_1)w(y_2)\mathcal{G}(y_1,t_1;y_2,t_2)\,,
\end{split}
\label{eq:probability}
\end{equation}
where we have introduced the fermionic correlator
\begin{equation}
\mathcal{G}(y_1,t_1;y_2,t_2)=\left\langle\hat\psi_R(y_1,t_1)\hat\psi^\dagger_R(y_2,t_2)\right\rangle_\Omega\,.
\label{eq:def_G}
\end{equation}
Using the identity in \eqref{eq:app:identity2} and introducing the shorthand notations
\begin{equation}
z_\eta=x-\eta ut\,,\qquad z^\eta_i = y_i-\eta u t_i \qquad (i=1,2)\,,
\end{equation}
the correlator $\cal{G}$ is expressed in terms of the bosonic Green function
\begin{equation}
\label{eq:G}
\begin{split}
G(\pm z) &= \langle \hat \phi_\mp(z)\hat \phi_\mp(0)\rangle_\Omega - \langle \hat \phi_\mp^2(0) \rangle_\Omega\\ & = \frac{1}{2\pi} \log \frac{a}{a\pm i z}
\end{split}
\end{equation}
as
\begin{equation}
\label{eq:Lambda}
{\cal G}(z_1^{\pm};z_2^{\pm})=\frac{1}{2\pi a}\; e^{2 \pi A_+^2 G(z_2^+-z_1^+)} \;e^{2\pi A_-^2 G(z_1^--z_2^-)}\,.
\end{equation}
The inverse lifetime $2\gamma$ is reated to the transition probability by
\begin{equation}
2\gamma=\lim_{t\to+\infty}\dot P_{1 \to 0}(t)\,.
\label{eq:lifetime}
\end{equation}
Performing the time derivative we obtained (see Appendix \ref{app:lifetime})
\begin{equation}
\label{eq:gamma}
\gamma =\gamma_0\, \frac{v_{\rm F}}{2 \pi} \int dk\; \mathcal{A}_R(k,\epsilon_0)\,  \left| \tilde \xi \left(k_0-k\right)\right|^2,
\end{equation}
with
\be 
\gamma_0 =\frac{|\lambda|^2}{2 v_{\rm F}}. 
\label{gamma0}
\ee
}
\begin{figure}[t]
	\centering
	\includegraphics[width=\linewidth]{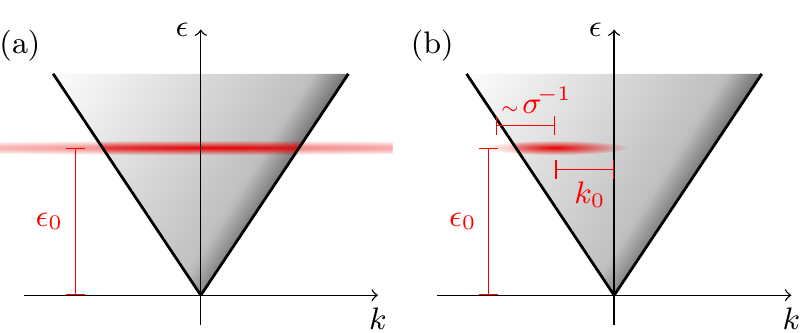}
	\caption{(Color online) Sketch of the overlap between the spectral function $\mathcal{A}_R(k,\epsilon)$ (in gray) and $|\tilde \xi(k)|^2$ (in red). The latter is represented with a horizontal line at energy $\epsilon_0$ since we are considering injection of an electron with well defined energy. Panel (a): local injection ($\sigma\to 0$).  Panel (b): non-local injection ($\sigma \sim 2 u \epsilon_0^{-1}$) with a finite extension in $k$ region for  $|\tilde \xi(k)|^2$ centered around $k_0$.}
	\label{fig:spectral1}
\end{figure}
Here, $\tilde \xi(k)$ is the Fourier transform of the real envelope function $\xi(y)$ (see Eq. \eqref{eq:app:xi_tilde}) and thus $|\tilde \xi(k_0-k)|^2$ is centered around $k=k_0$. The function 
\begin{equation}
\label{eq:spectral}
\begin{split}
\mathcal{A}_R(k,\epsilon>0)=&\; \frac{ 2 \pi \, e^{- \epsilon a/u}}{A_-^2\Gamma^2(A_-^2)}\; \left(\frac{a}{2 u}\right)^{2A_-^2}   \left(\epsilon+uk \right)^{A_-^2}  \\ & \left(\epsilon-uk\right)^{A_-^2-1} \theta(\epsilon-u|k|),
\end{split}
\end{equation}
is the spectral function of the right edge channel \cite{giamarchi2003quantum}. Recall that $k$ and $\epsilon$ are defined as momentum and energy with respect to $k_{\rm F}$ and $E_{\rm F}$ respectively. 
Equation \eqref{eq:gamma} has a clear physical interpretation: ${2}\gamma$ represents a tunneling rate and is proportional to the overlap between the  spectral function $\mathcal{A}_R(k,\epsilon_0)$ and the $k$ ``spectrum'' of the injected electron, described by $|\tilde \xi(k_0-k)|^2$. In Fig. \ref{fig:spectral1} one can see this overlap in the energy and momentum space. The region where $\mathcal{A}_R(k,\epsilon)\neq 0$ is filled in gray, showing that in the presence of e-e interactions the spectral function broadens and does not vanish away from the mass shell ($\epsilon = uk$). The injected electron has a well defined energy $\epsilon_0>0$ and thus the function $|\tilde \xi(k_0-k)|^2$ is represented as a red horizontal line at $\epsilon_0$, centered around $k=k_0$ with an extension of the order of $\sigma^{-1}$. Panel (a) refers to local injection: $\sigma \to 0$, with $|\tilde \xi (k_0-k)|^2\sim 1$. Here, the momentum $k_0$ is not relevant and the overlap is along the darker red line over the gray cone. In this limit the integral in Eq. \eqref{eq:gamma} can be solved analytically, giving the local rate 
\begin{equation}
\label{eq:gamma_local}
\gamma^{loc}= K \gamma_0 \; \frac{\left(K \bar a\right)^{2A_-^2}}{\Gamma(1+2A_-^2)}\;  e^{-K \bar a}\,,
\end{equation}
with $\bar a = a \epsilon_0 /v_{\rm F}$ the dimensionless cut-off.  Note that $\gamma_0$ in Eq. (\ref{gamma0})
represents the asymptotic value of $\gamma^{loc}<\gamma_0$ in the non interacting limit $K\to 1$. 

Fig. \ref{fig:spectral1}(b) shows a non-local injection. Here, $|\tilde \xi(k_0-k)|^2$
is centered around $k_0$, chosen in the figure to be negative, with a width $\sim\sigma^{-1}$.  The overlap between the two functions is significantly smaller with respect to (a) and it further reduces as long as $k_0$ is pushed away from the gray cone.  In addition,
for a given interaction strength $K$, and momentum $k_0$, the overlap decreases as $\sigma$ increases with the result  $\gamma<\gamma^{loc}< \gamma_0$.

\begin{figure}[t]
	\centering
	\includegraphics[width=.9\linewidth]{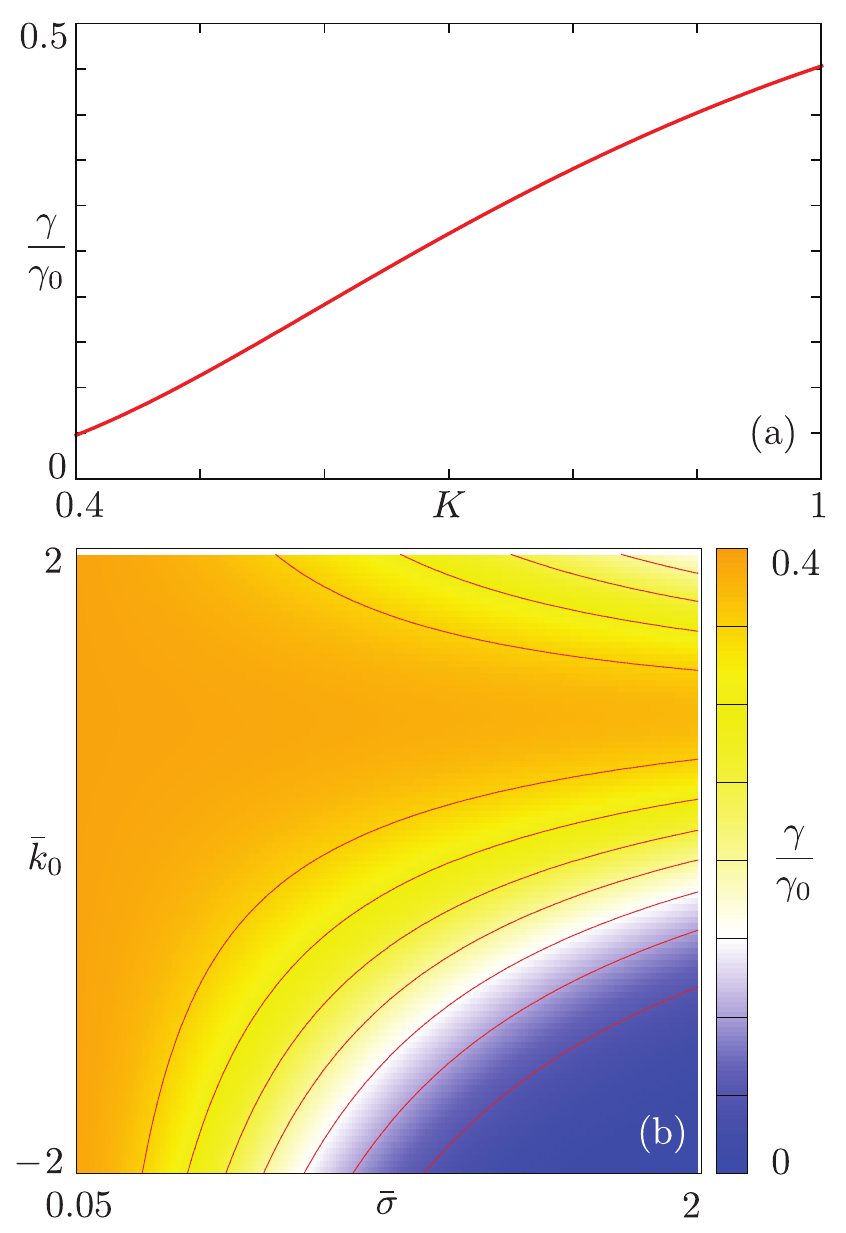}
	    \caption{(Color online) Panel (a): ratio $\gamma/\gamma_0$ as a function of interaction strength $K$ with $ \bar{\sigma}= 0.9$ and $ \bar{k}_0 = -1.2$. Panel (b): density
plot of $\gamma/\gamma_0$ as a function of $\bar{\sigma}$ ($x$-axis) and $\bar{k}_0$ ($y$-axis) with $K=0.6$. In both panels $\bar a = 1/40$.}
\label{fig:normalizzazione}
\end{figure}

In order to discuss quantitative results, we consider a gaussian envelope function with extension $\sigma$
\begin{equation}
	\label{eq:gaussian}
\xi(y) = \frac{1}{\sqrt{\pi} \sigma}\; e^{-y^2/\sigma^2}\,. 
\end{equation}
Note that for $\sigma\to 0$ we recover the point-like injection $\xi(y)=\delta(y)$, while increasing  $\sigma$ the injection extension increases with a decreasing amplitude. For convenience we introduce the dimensionless parameters
\begin{equation}
\label{eq:dimensionless}
\bar \sigma = \frac{\sigma \epsilon_0}{v_{\rm F}}, \quad \bar k_0 = \frac{k_0 v_{\rm F}}{\epsilon_0}.
\end{equation}

The dependence of the ratio $\gamma/\gamma_0$ on different parameters is reported in Fig. \ref{fig:normalizzazione}, where the relation $\gamma < \gamma_0$ clearly emerges. Panel (a) shows the suppression of the tunneling rate as the interaction strength increases, a well-known feature of HLL. Parameter $\bar k_0$, considered in panel (b), does not affect $\gamma$ as long as local-tunneling is concerned but becomes more and more relevant as $\bar \sigma$ increases. In particular, $\gamma$ significantly diminishes when $k_0$ is pushed away from the momentum range where the spectral function $\mathcal{A}_R(k,\epsilon_0)$ has finite values (see also Fig. \ref{fig:spectral1}).

\section{Charge density}
\label{sect:charge}
The above general method is applied here to study the time evolution of the charge density variation $\delta n(x,t)$, defined as in 
Eq. (\ref{eq:averageO}) with $\hat{O}\equiv \hat{n}$. Note that charge is measured in units of the electron's one so that charge density exactly equals particle density $\hat n(x,t)$. 
The latter can be expressed in terms of chiral bosonic fields as 
\begin{equation}
\label{eq:density}
\hat n(x,t) = - \sqrt{\frac{K}{2 \pi}} \sum_{\eta} \eta \partial_x \hat \phi_\eta\,.
\end{equation}

As shown in Appendix \ref{app:},  the average factor $\mathcal{I}_{O=n}$ in Eq. (\ref{eq:I}) can be evaluated yielding
\begin{equation}
\mathcal{I}_n = \sum_{\eta=\pm 1} q_\eta \left[\frac{1}{\pi} \frac{a}{a^2+(z_\eta-z^{\eta}_{2})^2}\right] \cal{G}\,,
\end{equation}
where
\be
q_\eta=\sqrt{K}A_\eta=\frac{1+\eta K}{2}
\ee
and $\mathcal{G}$ is given in \eqref{eq:def_G}.

The charge density is then expressed inserting $\mathcal{I}_n $ into the average (\ref{eq:averageO}).
It results into the sum of two chiral contributions $\delta n(x,t)=\sum_{\eta}\delta n_\eta(z_\eta)$, with
	\begin{equation}
	\begin{split}
	\label{eq:general_n}
	\delta n_\eta(z_\eta) =&\;  \frac{q_\eta |\lambda|^2}{\pi a} \;  \Re \int_0^t\!\!dt_2\!\!\int_0^{t_2}\!\!dt_1\!\!\iint_{-\infty}^{+\infty}\!\!dy_1dy_2\;\;\\
	& \qquad \Xi(t_1,t_2,y_1,y_2)\;   \delta (z_\eta - z^\eta_2)\;  \\ &\qquad e^{2 \pi A_+^2 G(z_2^+-z_1^+)} \;e^{2\pi A_-^2 G(z_1^--z_2^-)}. 
	\end{split}
	\end{equation}

\subsection{Charge fractionalization}

The total amount of injected charge that travels in a given direction ($\eta=\pm$) is
\begin{equation}
\label{eq:Qeta}
\mathcal{Q}_\eta = \int_{-\infty}^{+\infty} \!\!dx\; \delta n_\eta(x,t\to \infty).
\end{equation}
This integral can be easily performed from Eq. (\ref{eq:general_n}) for $\delta n_\eta(z_\eta)$. 
One finds $\mathcal{Q}_\eta=q_\eta \mathcal{Q}$ where 
\begin{equation}
\label{eq:Qstep1}
\mathcal{Q} = |\lambda|^2 \int_0^\infty\!\!dt_2\!\!\int_0^{t_2}\!\!dt_1\!\!\iint_{-\infty}^{+\infty}\!\!dy_1dy_2\; \left[\Xi\,{\cal G} + h.c.\right]
\end{equation}
represents the total amount of charge injected in the system.
{Note that the previous relation can be also written as
\begin{equation}
\label{eq:Qstep2}
\mathcal{Q} = |\lambda|^2 \iint_0^{+\infty}\!\!dt_1dt_2\!\!\iint_{-\infty}^{+\infty}\!\!dy_1dy_2\; \Xi\, {\cal G}\,.
\end{equation}}
We thus recover the well-known \cite{maslov1995landauer,safi1995transport} expression for charge fractionalization factors
\begin{equation}
\label{eq:q_eta}
\frac{\mathcal{Q}_\eta}{\mathcal{Q}_++\mathcal{Q}_-} = q_\eta = \frac{1+\eta K}{2}
\end{equation}
that depend {\em only} on the interaction strength $K$.
As discussed in Appendix \ref{app:} all contributions to $\mathcal{Q}_\eta$ are due to ${\cal I}_n^{(a)}$ and not to the polarization term ${\cal I}_n^{(b)}$.
 
For $t\gg1/(2\gamma)$ the QD level is empty and the total amount of injected charge $\mathcal{Q}=\mathcal{Q}_++\mathcal{Q}_-$ {is expected to} satisfy $\mathcal{Q}=1$.
{It is indeed shown in Appendix \ref{app:q} that, as long as $\gamma\ll\epsilon_0$, the condition $\mathcal{Q}=1$ holds.}

\subsection{Charge density profile after local injection}
We now focus on the local-injection limit $\xi(y) = \delta(y)$, in order to study interactions effects on the charge density profile. Integrating Eq. (\ref{eq:general_n}) one has 
\begin{equation}
\label{eq:n_local_step1}
\begin{split}
\delta n_\eta&(x,t) = \frac{q_\eta |\lambda|^2}{2 \pi a u} \;  2 \Re \int_0^t\!\!dt_2\!\!\int_0^{t_2}\!\!dt_1\; \beta^*(t_1)\beta(t_2)\\
&\qquad \delta(t_2-t-\frac{\eta x}{u}) \left(\frac{a}{a+iu(t_1-t_2)}\right)^{1+2A_-^2}.
\end{split}
\end{equation}
We observe that, apart from the fractionalization factors $q_\eta$, the two chiral charge density packets share the same mirrored shape
\begin{equation}
\label{eq:n+=n-}
\frac{\delta n_+(x,t)}{q_+} = \frac{\delta n_-(-x,t)}{q_-}.
\end{equation}
As a consequence, we can focus only on the right-moving packet ($\eta=+$). We analyze the corresponding charge current $j_{+}(\tau) = u \delta n_+(\tau)$ with $\tau = t-x_D/u$, 
flowing through a ``detection'' point $x_D>0$ away from the injection region. The integral over $t_2$ in Eq. (\ref{eq:n_local_step1}) can be easily performed yielding
\begin{equation}
\label{eq:j_local}
j_{+}(\tau) = 2 q_+  \gamma_0 \; \theta(\tau) \; \exp\left[-2 \tau \gamma\right] \;  \Re\left[ C_1(\tau)\right]~,
\end{equation}
where ($m \in \mathbb{N}$)
\begin{equation}
\label{eq:C_n}
C_m(\tau) = \frac{\epsilon_0}{\pi \bar a^m} \int_{-\tau}^0 ds\; e^{-s\gamma} e^{i s \epsilon_0} \left(\frac{\bar{a}}{\bar a + i s \epsilon_0 K^{-1}}\right)^{m+2A_-^2}\,.
\end{equation}

First of all we note that, because of causality, $j_{+}(\tau)\neq 0$ only for $\tau>0$, since an excitation created in $x=0$ takes exactly a time $x_D/u$ to reach the detection point.  Another clear feature is the exponential decrease $e^{-2 \gamma \tau }$ due to the QD single level inverse lifetime ($2\gamma$). The presence of the interacting helical Fermi sea is taken into account by the function $C_1(\tau)$ \cite{footnote:C_n}.
\begin{figure}[t]
	\centering
	\includegraphics[width=\linewidth]{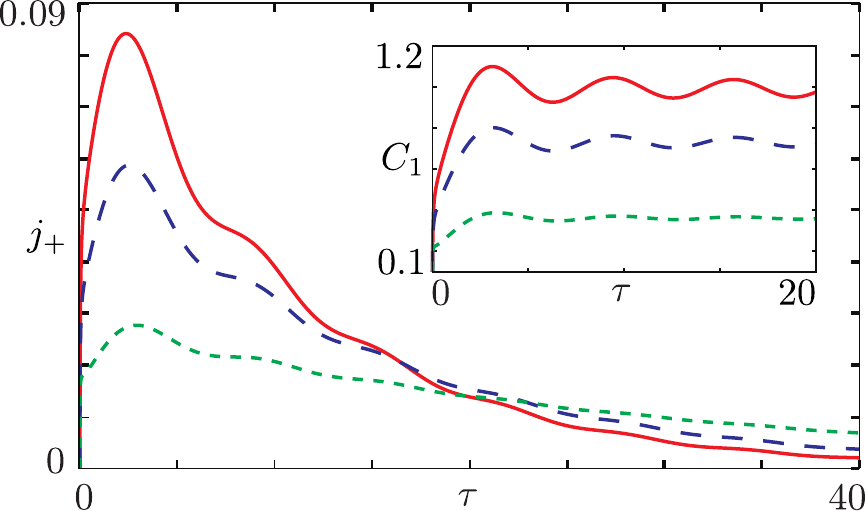}
	\caption{(Color online) Charge current $j_{+}(\tau)$ (in units of $\epsilon_0$) flowing in the right direction through the detection point $x_D>0$ as a function of time (in units of $\epsilon_0^{-1}$). Different interaction strengths are considered: solid red line $K=1$ (non-interacting case), dashed blue $K=0.8$, and green dotted $K=0.6$. The inset shows the function $C_1(\tau)$ with the same color coding. Parameters: $\gamma_0=0.05\;  \epsilon_0$ and $\bar a = 1/40$.}
	\label{fig:charge_plot}
\end{figure}
Fig. \ref{fig:charge_plot} shows all these features. The decreasing exponential behavior is clearly visible as well as the increase of the QD level lifetime $(2\gamma)^{-1}$ as interactions strength increases. Function $C_1(\tau)$, plotted in the inset, is characterized by a global decrease, while increasing interaction strength.  
It also presents oscillations with a period given by $2 \pi \epsilon_0^{-1}$ and an amplitude damped by interactions. This fact is due to the smearing of the Fermi function, which weakens the effects of the Fermi sea.

Similar qualitative features are expected in the case of non-local injection, where however the pulse will be less localized. Bigger effects related to the nature of the injection process manifest at the level of energy partitioning and therefore will be discussed more in detail later.
{Although challenging, experimental detection of such fractional charge packets could be performed. High resolution time-resolved measurements are indeed possible in quantum Hall bars, using a quantum point contact as a shutter on the ps scale~\cite{kataoka13, kataoka16} that allows the study of charge packet profiles~\cite{kamata2014fractionalized}. Different measurement schemes, based on Hong-Ou-Mandel interferometry~\cite{fevenatcomm, fevephyse}, have also been used to detect charge profiles.} 

\section{Energy density}
\label{sec:energy}

The injected electron transfers into  the helical edge not only charge but also energy. We then start focusing on the evaluation of the energy density (see Eq. \eqref{eq:def_H}) variation, proceeding  along the lines discussed in the previous Section. Considering $\mathcal{I}_{O=\mathcal{H}}$ in Eq. \eqref{eq:I} and the commutator relation in Eq. \eqref{eq:app:commutator} one can derive the following expression
\begin{equation}
\label{eq:IH}
\begin{split}
&\mathcal{I}_\mathcal{H} = \sum_{\eta} \frac{u}{2}
\meanOmega{\hat \psi_R(y_1,t_1) \left[:\!\left(\hat \partial_x \phi_\eta(z_\eta)\right)^2\!\!:\,,\, \hat \psi_R^\dagger(y_2,t_2)\right]} \\ & = -\!
\sum_{\eta} \!\frac{u \eta  A_\eta\sqrt{ \pi}}{\sqrt{2}}\!\left(\frac{1}{\pi}\frac{a}{a^2+(z_\eta-z_2^\eta)^2}\right) \partial_x \!\left(\!\mathcal{M}_\eta^{(a)}\! +\! \mathcal{M}_\eta^{(b)}\right)
\end{split}
\end{equation}
with
\begin{align}
\mathcal{M}_\eta^{(a)} &= \meanOmega{\hat \psi_R(y_1,t_1)\, \hat \phi_\eta(z_\eta) \,\hat \psi_R^\dagger(y_2,t_2)}\\
\mathcal{M}_\eta^{(b)} &= \meanOmega{\hat \psi_R(y_1,t_1) \,\hat \psi_R^\dagger(y_2,t_2)\, \hat \phi_\eta(z_\eta) }.
\end{align}
These average functions are evaluated in Appendix \ref{app:energy} with the final result
\begin{equation}
\label{eq:IH_step2}
\begin{split}
\mathcal{I}_\mathcal{H} &= u \sum_{\eta} A_\eta^2\, \mathcal{G} \, \left[ i \frac{\eta}{2} \partial_{z_2^\eta}\left(\frac{1}{\pi} \frac{a}{a^2 + (z_\eta-z_2^\eta)^2}\right)\right.\\ &
\qquad \left. + \left(\frac{1}{\pi} \frac{a}{a^2 + (z_\eta-z_2^\eta)^2}\right) \frac{1}{a+i\eta (z_\eta- z_1^\eta )}\right].
\end{split}
\end{equation}
This formula allows to express the total energy density profile $\delta\mathcal{{H}}(x,t)$ in Eq. \eqref{eq:averageO} as a sum of the 
left and right moving contributions $\delta\mathcal{{H}}(x,t) =\sum_\eta \delta\mathcal{H}_\eta(z_\eta)$, with
\begin{equation}
\begin{split}
\label{eq:deltaH}
\delta \mathcal{H}_\eta (z_\eta) &= \frac{uA_\eta^2 |\lambda|^2}{\pi a^2} \;\Re \int_0^t\!\!dt_2\!\!\int_0^{t_2}\!\!dt_1\!\!\iint_{-\infty}^{+\infty}\!\!dy_1dy_2 \;\;\Xi\,\mathcal{G}\\
& \qquad \left(\frac{a}{a+i\eta (z_\eta-z_1^\eta)} + i\frac{\eta a}{2} \partial_{z_2^\eta}\right) \delta(z_\eta - z_2^\eta). 	
\end{split}
\end{equation}

\subsection{Energy density profile after local-injection}
In order to highlight the effects of e-e interactions we start by discussing the local-injection limit.  
Integrating Eq. (\ref{eq:deltaH}) over space with $\xi(y)=\delta(y)$ one obtains 
\begin{equation}
\begin{split}
&\delta\mathcal{H}_\eta(z_\eta) = \frac{A_\eta^2 |\lambda|^2}{\pi a^2}\;  \Re \int_0^t\!\!dt_2\!\!\int_0^{t_2}\!\!dt_1 \;\;\beta^*(t_1)\beta(t_2) \\&\qquad \left[\frac{a}{2ui} \left(\frac{a}{a+iu (t_1-t_2)}\right)^{2A_-^2+1} \partial_{t_2} \delta(t_2-t+\frac{\eta x}{u})\right. \\
& \left. \qquad + \left(\frac{a}{a+iu (t_1-t_2)}\right)^{2A_-^2+2} \delta(t_2-t+\frac{\eta x}{u})  \right].
\label{penergy}
\end{split}
\end{equation}
Similarly to charge, the two chiral energy density packets share the same mirrored shape as long as local-injection is concerned
\begin{equation}
\frac{\delta\mathcal{H}_+(x,t)}{A_+^2} = 
\frac{\delta\mathcal{H}_-(-x,t)}{A_-^2}.
\end{equation}
We then focus on the right moving energy packet ($\eta=+$), by analyzing the instantaneous energy power $P_{+}(\tau)=u\,  \delta\mathcal{H}_+(\tau)$ that flows through the  ``detection'' point $x_D$. Integration of  (\ref{penergy}) over $t_2$ leads to $(\tau=t-x_D/u)$
\begin{equation}
\begin{split}
\label{eq:Ptau}
P_{+}(\tau) =  A_+^2 \, \gamma_0  \, \epsilon_0 \; \theta(\tau) \exp\left[-2 \tau \gamma\right]\, \Re\left[C_\mathcal{H}(\tau)\right]~,
\end{split}
\end{equation}
with
\begin{equation}
C_\mathcal{H}(\tau) = \frac{\epsilon_0 - i \gamma}{\epsilon_0} C_1(\tau) + \frac{1}{K} \left(1-2A_-^2\right) C_2(\tau)
\end{equation}
and $C_m(\tau)$ ($m=1,2$) given in Eq. \eqref{eq:C_n}. 
\begin{figure}[t]
	\centering
	\includegraphics[width=\linewidth]{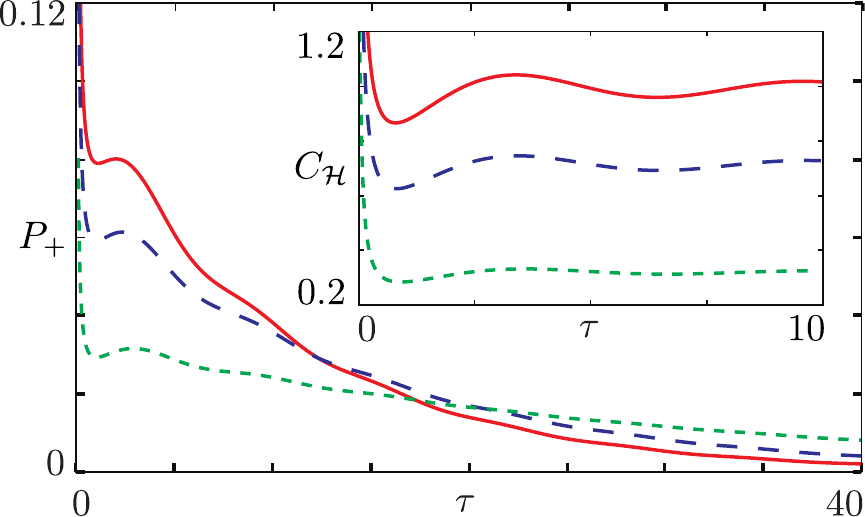}
	\caption{(Color online) Instant energy power $P_{+}(\tau)$ (in units of $\epsilon_0^2$) flowing through the detection point $x_D>0$ as a function of time $\tau$ (in units of $\epsilon_0^{-1}$). Different interaction strengths are considered: solid red line ($K=1$), dashed blue $K=0.8$, and green dotted $K=0.6$. Inset: function $C_\mathcal{H}(\tau)$ with the same color code for interactions. Parameters: $\gamma_0=0.05\;  \epsilon_0$ and $\bar a = 1/40$.}
	\label{fig:energy_plot}
\end{figure}
In Fig. \ref{fig:energy_plot} the instantaneous energy power $P_+(\tau)$ is plotted as a function of time for different interaction strength.
As for charge current, it reflects  causality, ensured by $\theta(\tau)$, and  the exponential decay related to the QD level inverse lifetime $2 \gamma$, with analogous behaviors.
The function $C_\mathcal{H}(\tau)$ (plotted in the inset) features also a spike at $\tau =0$, even in the non-interacting case, reflecting the sudden turning on of the injection process and the consequent excitation, at short times, of energy modes, even higher than $\epsilon_0$.

 \subsection{Energy partitioning}
To analyze energy partitioning phenomena, we now focus on the total amount of energy  that travels in a given direction once the injection is concluded
\begin{equation}
\label{eq:E__eta}
E_\eta = \int_{-\infty}^{+\infty}\!\!dx\; \delta\mathcal{H}_\eta(x,t \to \infty)~.
\end{equation}
Using the expression \eqref{eq:deltaH} for $\delta\mathcal{H}_\eta(x,t)$ one has
\begin{equation}
\label{eq:E_eta_step1}
\begin{split}
E_\eta &= \frac{u A_\eta^2|\lambda|^2}{2\pi a^2}  \int_0^{+\infty}\!\!dt_2\!\!\int_0^{t_2}\!\!dt_1\!\!\iint_{-\infty}^{+\infty}\!\!dy_1dy_2 \;\\
& \qquad \left[ \Xi \;\; e^{2 \pi G(z_2^+-z_1^+) g^+_\eta} e^{2 \pi G(z_1^--z_2^-) g^-_\eta} + h.c. \right]~,
\end{split}
\end{equation}
where $g_\eta^\pm = A_\pm^2 + (1\pm\eta)/2$.
In passing we note that the term $I_{\mathcal{H}}^{(b)}$, present  in Eq. \eqref{eq:I}, does not contribute to this integrated quantity \cite{footnote:H}.
The above expression can be conveniently represented in Fourier space (similarly to what {has been done} in Appendix \ref{app:lifetime}) as
\begin{equation}
\begin{split}
\label{eq:Eeta}
E_\eta  &= \frac{K A_\eta^2 \gamma_0}{2\pi} \left(\frac{K \bar a}{2 \epsilon_0}\right)^{2A_-^2} \!\! \frac{1}{\Gamma(g_\eta^-)\Gamma(g_\eta^+)}
\; \; \int_0^{+\infty}\!\! d\epsilon_+ \\ & \qquad \left|\tilde \beta(\epsilon_+)\right|^2 e^{-K \bar a \frac{\epsilon_+}{\epsilon_0}} \!\int_{-\epsilon_+}^{+\epsilon_+}\!\!d\epsilon_- (\epsilon_++\epsilon_-)^{g_\eta^+-1}\\&\qquad   (\epsilon_+-\epsilon_-)^{g_\eta^--1} \left|\tilde \xi\left(k_0 - \epsilon_-/u\right)\right|^2~.
\end{split} 
\end{equation}
The key quantities to discuss are  the energy partitioning factors defined as  
 \begin{equation}
 \label{p_fraction}
 p_\eta = \frac{E_\eta}{E_+ + E_-}~.
 \end{equation}
They indeed represent the fraction of the total energy $E=E_++E_-$ that propagates in the  direction $\eta=\pm$.  Concerning the total contribution $E= E_+ + E_-$, we demonstrate in Appendix \ref{app:ploc} that $E = \epsilon_0$ as long as $\gamma \ll\epsilon_0$.

In the local injection limit  $\tilde \xi(k)=1$ one has (see Appendix \ref{app:ploc})
\begin{equation}
\label{eq:p_loc}
p_\eta^{loc} = \frac{A_\eta^2}{A_-^2+A_+^2} = \frac{(1+\eta K)^2}{2(K^2+1)}.
\end{equation}
Namely, energy partitioning has a ``universal'' character, i.e. $p_\eta^{loc}$ does not depend on injection parameters but {\em only} on interaction strength, in agreement with the partitioning of DC energy transport found in Ref. \onlinecite{karzig2011energypart}. We have thus shown that this ``universal'' feature still holds also in the case of time-resolved single electron injection from a mesoscopic capacitor. 

\begin{figure}
    \centering
    \includegraphics[width=\linewidth]{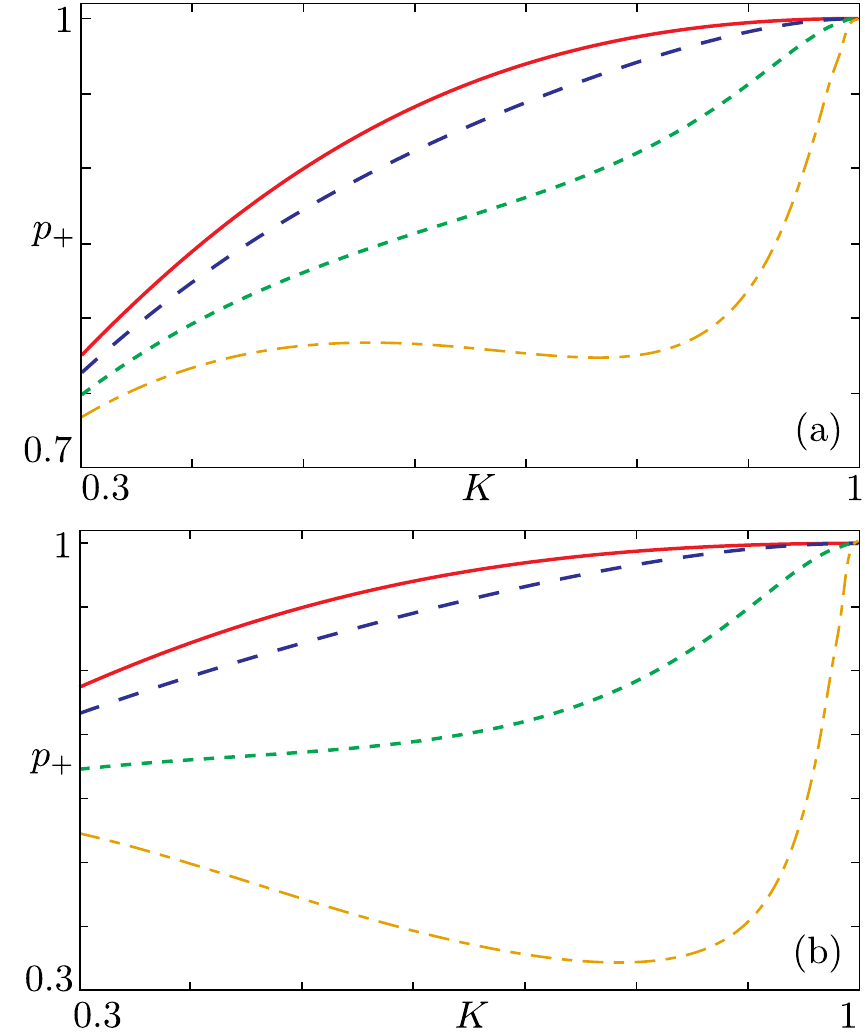}
    \caption{(Color online) Energy partitioning factor $p_+$ as a function of the interaction strength $K$. In panel (a) $\bar{k}_0 = 0$, with $\bar \sigma \to 0$ (solid red), $\bar \sigma = 2$  (dashed blue), $\bar \sigma = 3$ (dotted green) and $\bar \sigma = 3.75$ (dot dashed orange). Panel (b) shows $\bar{k}_0=-1.2$ with 
$\sigma \to 0$ (solid red), $\bar \sigma = 0.9$ (dashed blue), $\bar \sigma = 1.5$ (dotted green) and $\bar \sigma = 1.95$ (dot dashed orange). Parameters:
$\gamma_0 = 0.05\,\epsilon_0$ and $\bar a = 1/40$.}
    \label{fig:luttinger}
\end{figure}
On the other hand, it can be shown that such universality breaks down as the tunneling region increases. 
In order to quantitatively highlight this deviation we present below results for the right moving energy fraction $p_+$ in  Eq. (\ref{p_fraction}), using the gaussian envelope $\xi(y)$ (see Eq. \eqref{eq:gaussian}). 

\begin{figure}
	\centering
	\includegraphics[width=0.8\linewidth]{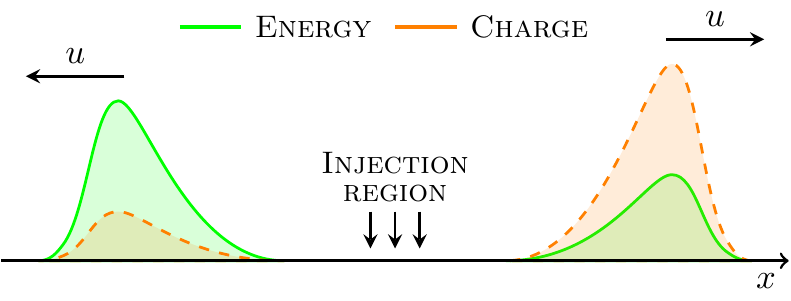}
	\caption{(Color online) Cartoon showing the strong direction separation of energy (solid green) and charge (dashed orange) for $K=0.8$, $\bar{\sigma}=1.95$ and $\bar{k}_0=-1.2$. The majority of charge ($80\%$) travels to the right while most of the energy (about $65\%$) moves to the left.}
	\label{fig:Cartoon}
\end{figure}

Fig. \ref{fig:luttinger} shows two representative cases of energy partitioning as a function of interaction strength.  
The ``universal'' limit $p_+^{loc}$ (\ref{eq:p_loc}) is drawn with a solid red line. Panel (a), has $\bar k_0 = 0$, and shows deviations from the universal limit as $\bar \sigma$ increases, with $0.5<p_+(K)<p_+^{loc}(K)$. These deviations are even more striking for negative values of $k_0$ as shown in panel (b) with $\bar k_0 = -1.2$. Here, it is even possible to achieve $p_+(K)<0.5$ for a wide range of interaction strength (dot-dashed curve). This means that, due to interactions and non local tunneling, an electron injected into the right branch, creates an energy packet that mostly travels to the left, while the charge still continues to move mainly to the right ($q_+>q_-$). Fig. \ref{fig:Cartoon} represents the cartoon of this opposite charge and energy propagation.  
\begin{figure}
      \centering
            \includegraphics[width=\linewidth]{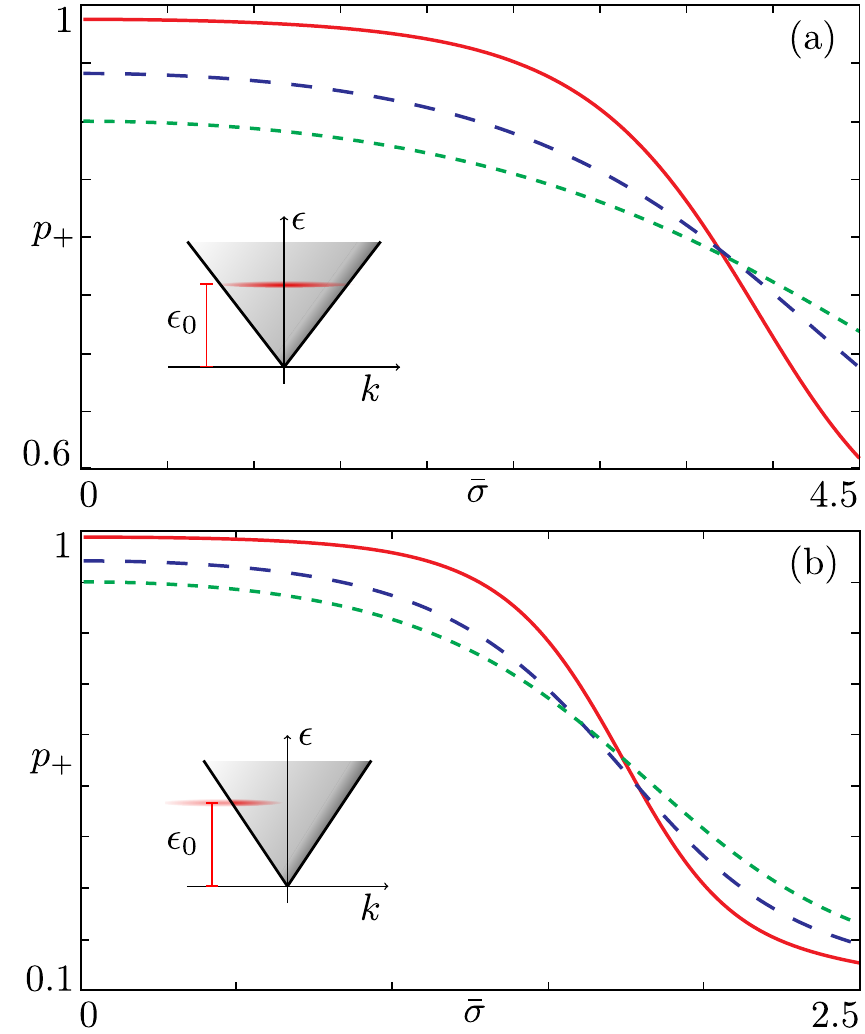}
      \caption{(Color online) Energy partitioning factor $p_+$ as a function of the tunneling region width $\bar{\sigma}$.
Each line refers to different  interaction parameter: $K=0.8$ (solid red), $K=0.6$ (dashed blue) and $K=0.5$ (dotted green). In panel (a) $ \bar{k}_0=0$
while panel (b) $ \bar{k}_0=-1.2$.  The insets show the overlap, at the same interaction strength, between the edge spectral function (in gray) and the
momentum ``spectrum'' of the injected electron (in red), along the lines of Fig. \ref{fig:spectral1}. The momentum  $ \bar{k}_0$ is the same of the hosting
panel. Parameters: $\gamma_0 = 0.05 \,\epsilon_0$ and $\bar a = 1/40$.}
      \label{fig:sigma}
\end{figure}
To clarify the physical interpretation of this effect, we consider in Fig. \ref{fig:sigma} the energy partitioning factor $p_+$ as a function of $\bar \sigma$ for different interaction strength. In panel (a) $\bar k_0=0$ while in panel (b) $\bar k_0=-1.2$. For $\bar \sigma \to 0$ one recovers the ``universal'' behavior, while deviations from it become relevant as $\bar \sigma$ increases and reaches $\bar \sigma \gtrsim 1$. Comparing the two panels, note that these deviations emerge at smaller $\bar\sigma$ when $\bar k_0$ is significantly different from $\bar k_0=0$. This fact can be understood considering again the overlap between the spectral function $\mathcal{A}_R(\epsilon,k)$ and the injected electron momentum ``spectrum'' $|\tilde \xi(k_0-k)|^2$ represented as insets of the two main panels in  Fig. \ref{fig:sigma}. Here, we sketched two typical situations with the same interaction and momentum $\bar{k}_0$ as given in the main panel. Non-universal effects appear only when the red line does not cover the whole gray region, whose extension at $\epsilon=\epsilon_0$ is given by $2 \epsilon_0 K/v_{\rm F}$ (see Eq. \eqref{eq:spectral}).
Therefore, if one considers $\bar k_0 = 0$ (panel (a)) it is necessary $\bar \sigma \gtrsim K^{-1}$ in order to brake the energy partitioning universality. By contrast, for a negative $\bar k_0 = -1.2$ (panel (b)), a smaller $\bar \sigma$ will be required since the overlap is already smaller.

Note that all these deviations are much less pronounced (and then not shown) for $\bar k_0>0$ since even with extended tunneling, the transferred momentum lies near the right electron branch, leading to $p_+(K)>p_+^{loc}$.  As a last comment, the non-interacting limit $K\to1$ shows always $p_+=1$, regardless of all the other parameters. Energy partitioning is indeed a manifestation of e-e interactions and so, if they're absent, all the energy added to the system after an $R$-electron injection goes to the right.

In the end, we want to stress that the condition $\bar \sigma \sim 1$, although challenging, is consistent with the boundary imposed by the setup. Non-universal features of energy partitioning can thus play an important role when a non-local injection is concerned. In particular, it is possible to directly control the energy flow after a single electron injection, being able even to invert its direction with respect to the charge flow. {The energy flow, and its partitioning, could be inspected by means of nanocalorimetric measurements~\cite{giazotto12, pekola15}. Their implementation within a time-resolved detection scheme, analog to the time-dependent charge measurements~\cite{kamata2014fractionalized}, should also allow the study of the energy packet power profile $P_+(t)$. Additional information on fractional excitations can be obtained by measuring their energy distribution~\cite{karzig2011energypart, battistaprb12, julienne14} which, for example, can be probed with a QD detector acting as an energy filter~\cite{altimiras09}.}

\section{Conclusions}
In this work we have investigated the injection process of a single electron from a mesoscopic capacitor into the counterpropagating edge states of a topological insulator. Particular attention has been devoted to the role played by e-e interactions and how their presence affects the dynamics of both charge and energy density.
We have presented a time-dependent density matrix formalism to evaluate their time evolution after a single electron injection.
The  charge and energy profiles have been analyzed in presence of local and non-local tunneling.  Fractionalization phenomena, due to interactions, have been discussed, elucidating the differences between charge and energy. We have found that the latter is strongly affected not only by interactions but also by the nature of the tunneling process itself.
Indeed, we have shown that in presence of non local tunneling from a mesoscopic capacitor, it is possible to have situations in which charge and energy profiles flow in opposite directions and are completely decoupled. These results shed new lights on the single electron injection into an interacting system, with relevant implications for the field of electron quantum optics.

\begin{acknowledgments}
We acknowledge the support of the MIUR-FIRB2012 - Project HybridNanoDev (Grant  No.RBFR1236VV), EU FP7/2007-2013 under REA grant agreement no 630925 -- COHEAT, MIUR-FIRB2013 -- Project Coca (Grant No.~RBFR1379UX) and the COST Action MP1209.
\end{acknowledgments}

\appendix
\section{Calculation of the inverse lifetime}
\label{app:lifetime}
{In this Appendix we explicitly calculate the inverse lifetime $2\gamma$ defined in Eq. \eqref{eq:lifetime}.
Let us start from the result in Eq. \eqref{eq:probability}, which can be rewritten in the following form
\begin{equation}
\begin{split}
P_{1\to 0}(t)&=2|\lambda|^2\Re\int_0^t dt_2\int_0^{t_2}dt_1\iint_{-\infty}^{+\infty}dy_1 dy_2\,e^{i\epsilon_0(t_1-t_2)}\\
		&\quad w^*(y_1)w(y_2)\mathcal{G}(y_1,t_1;y_2,t_2)\,.
\end{split}
\end{equation}
It is now straightforward to perform the time derivative, obtaining
\begin{equation}
\begin{split}
\dot P_{1 \to 0}(t)&=2|\lambda|^2\Re\int_0^{t}dt_1\iint_{-\infty}^{+\infty}dy_1 dy_2\,e^{i\epsilon_0(t_1-t)}\\
				 &\quad w^*(y_1)w(y_2)\mathcal{G}(y_1,t_1;y_2,t)\,.
\end{split}
\label{eq:pdot}
\end{equation}
We now express this quantity in Fourier representation. First, considering\cite{vannucciqh} 
\begin{equation}
\label{eq:app:trasf_fourier_exp}
e^{2 \pi g G(z)} = \frac{1}{\Gamma(g)}\left(\frac{a}{u}\right)^{g} \int_0^{+\infty} \!\!dE \; E^{g-1} e^{-i\frac{E z}{u}} e^{-\frac{Ea}{u}}\,,
\end{equation}
the fermionic function $\cal G$ in Eq.~(\ref{eq:pdot}) becomes
\begin{equation}
\begin{split}
{\cal G} &= \frac{1}{2\pi a}\frac{1}{\Gamma(A_-^2)\Gamma(A_+^2)} \left(\frac{a}{u}\right)^{1+2A_-^2}  \iint_0^{+\infty} \!\!dE_1dE_2  \\ &\qquad E_1^{A_-^2}E_2^{A_-^2-1} e^{-a \frac{E_1+E_2}{u}} e^{- i t_1 (E_1+E_2) }\\ &\qquad e^{ i t (E_1+E_2) }e^{- i y_1 \frac{E_2-E_1}{u}} e^{i y_2 \frac{E_2-E_1}{u}}.
\end{split}
\label{eq:app:trasf_G}
\end{equation}
Then, we introduce the Fourier transform of $w(y)$
\begin{equation}
\label{eq:app:xi_tilde}
\begin{split}
\tilde w\left(k\right) &= \int_{-\infty}^{+\infty}\!\!dy \; w(y) \, e^{i k y} = \\ &= \int_{-\infty}^{+\infty}\!\!dy \; \xi (y) \, e^{i y (k+k_0)} =  \tilde \xi (k+k_0)~.
\end{split}
\end{equation}
Using \eqref{eq:app:trasf_G} and \eqref{eq:app:xi_tilde} in \eqref{eq:pdot} we obtain
\begin{equation}
\begin{split}
\dot P_{1 \to 0}(t)&=\frac{|\lambda|^2}{\pi u}\frac{1}{\Gamma(A_+^2)\Gamma(A_-^2)}\left(\frac{a}{2u}\right)^{2A_-^2}\int_0^{+\infty}d\epsilon\int_{-\epsilon}^{+\epsilon}dE\\
				 &\quad \left|\tilde w(-Eu^{-1})\right|^2(\epsilon+E)^{A_-^2}(\epsilon-E)^{A_-^2-1}e^{-\frac{\epsilon a}{u}}\\
				 &\quad\Re\int_0^t ds\,e^{-i(\epsilon_0-\epsilon)s}\,.
\end{split}
\end{equation}
Recalling the definition \eqref{eq:lifetime} and using
\begin{equation}
\Re\int_0^{+\infty}ds\,e^{-i(\epsilon_0-\epsilon)s}=\pi\delta(\epsilon_0-\epsilon)\,,
\end{equation}
we find
\begin{equation}
\begin{split}
\gamma&=\frac{\gamma_0K}{A_-^2\Gamma^2(A_-^2)}e^{-\frac{\epsilon_0a}{u}}\left(\frac{a\epsilon_0}{2u}\right)^{2A_-^2}\\
			&\int_{-1}^{+1}d\chi\left|\tilde w\left(-\frac{\epsilon_0\chi}{u}\right)\right|^2(1+\chi)^{A_-^2}(1-\chi)^{A_-^2-1}\,,
\end{split}
\label{eq:app:gamma_general}
\end{equation}
with $\gamma_0=|\lambda|^2/(2v_{\rm F})$. Thus Eq. \eqref{eq:gamma} is proved, using the expression in Eq.~(\ref{eq:spectral}) for the spectral function and Eq.~(\ref{eq:app:xi_tilde}). Note that when $\tilde\xi(k)=1$ (local tunneling), the above
integral can be evaluated analytically, yielding
\begin{equation}
\begin{split}
&\int_{-1}^{+1} \!\!d\chi \; (1+\chi)^{A_-^2}(1-\chi)^{A_-^2-1}  = \\
& \qquad = 2^{2A_-^2} \int_{0}^{1} dx \frac{x^{A_-^2}}{(1-x)^{1-A_-^2}} = 2^{2A_-^2} \frac{A_-^2 \Gamma^2(A_-^2)}{\Gamma(1+2A_-^2)}\nonumber\,.
\end{split}
\end{equation}
This result leads to Eq. \eqref{eq:gamma_local} which holds in the case of local injection with $\sigma \to 0$.
}
\section{Calculation of ${\cal I}_n$}

\label{app:}
This Appendix is devoted to the evaluation of the average function 
\be
\label{a1}
\mathcal{I}_n= {\cal I}_n^{(a)}-{\cal I}_n^{(b)}= \meanOmega{\hat  \psi_R(y_1,t_1) \left[\hat n(x,t),\hat \psi_R^\dagger(y_2,t_2)\right]}~,
\ee
defined in Eq.~(15) and necessary in order to compute the density variation $\delta n(x,t)$ in Eq. (\ref{eq:averageO}). Let us start with the commutator in (\ref{a1}), which can be written in terms of chiral fields as 
\begin{equation}
\label{eq:density1}
\left[\hat n(x,t),\hat \psi_R^\dagger(y_2,t_2)\right]= -\sqrt{\frac{K}{2 \pi}} \sum_{\eta} \eta \left[\partial_x \hat \phi_\eta(z_{\eta}),\hat \psi_R^\dagger(y_2,t_2)\right]~,
\end{equation}
with $z_{\eta}=x-\eta ut$. Using the bosonized expression (\ref{eq:boson_eq}) with (\ref{eq:r_to_eta}) one has
\begin{equation}
\begin{split}
&\left[\partial_x \hat \phi_\eta(z_\eta),\hat \psi_R^\dagger(y_2,t_2)\right] = \\&\qquad 
= \frac{1}{\sqrt{2 \pi a}} \left[\partial_x  \hat \phi_\eta(z_\eta),e^{i\sqrt{2 \pi}\left(A_+ \hat \phi_+(z_2^+) + A_- \hat \phi_-(z_2^-)\right)}\right]
\end{split}
\end{equation}
with the boson fields satisfying c-number commutation relations \cite{vondelft, miranda2003}
\begin{equation}
\label{eq:bosonic_fields_commutator}
\left[\partial_x \hat \phi_\eta(x), \hat \phi_{\eta'}(y)\right] = i \eta\;  \delta_{\eta,\eta'}\,  \frac{1}{\pi}\frac{a}{a^2+(x-y)^2}\,.
\end{equation}
This allows to use the Baker Hausdorff relation \cite{vondelft} among two operators $\hat A$ and $\hat B$ (with a c-number commutator) $\left[\hat A, e^{\hat B}\right] = \left[\hat A,\hat B\right] e^{\hat B}$, arriving to 
\begin{equation}
\label{eq:app:commutator}
\begin{split}
&\left[\partial_x \hat \phi_\eta(z_\eta),\hat \psi_R^\dagger(y_2,t_2)\right] = \\&\qquad = - \eta  A_\eta\sqrt{2 \pi}\left(\frac{1}{\pi}\frac{a}{a^2+(z_\eta-z_2^\eta)^2}\right) \hat \psi_R^\dagger(y_2,t_2)~.
\end{split}
\end{equation}
Then, using Eq. (\ref{eq:density}),  we arrive at
\begin{equation}
\label{eq:commutatore}
\begin{split}
&\left[\hat n(x,t) , \hat \psi_R^\dagger(y_2,t_2)\right] =\\&\qquad = \sum_\eta q_\eta \left[\frac{1}{\pi} \frac{a}{a^2+(z_\eta-z^{\eta}_{2})^2}\right] \hat \psi_R^\dagger(y_2,t_2)\,.
\end{split}
\end{equation}
The average function $\mathcal{I}_n$ in Eq. (\ref{a1}) is then given by
\begin{equation}
\mathcal{I}_n\!=\!\sum_{\eta=\pm 1} q_\eta \left[\frac{1}{\pi} \frac{a}{a^2+(z_\eta-z^{\eta}_{2})^2}\right] 
\langle\hat \psi_R(y_1,t_1) \hat \psi_R^\dagger(y_2,t_2) \rangle_\Omega
\nonumber
\end{equation}

As a final step the fermionic Green function 
${\cal G}=\langle\hat \psi_R(y_1,t_1) \hat \psi_R^\dagger(y_2,t_2) \rangle_\Omega$, is expressed using the identity \cite{vondelft}
\begin{equation}
	\label{eq:app:identity2}
	\meanOmega{e^{-i \alpha \hat \phi_\eta(x)}  e^{i \alpha \hat \phi_\eta(y)}} = \exp\left[{\alpha^2 G(-\eta(x-y))}\right]~,
\end{equation}
with $G$ the bosonic Green function defined in Eq. \eqref{eq:G}. 
In writing Eq. \eqref{eq:general_n}, as long as $a$ is the smallest length scale, it is possible to approximate 
\begin{equation}
\label{eq:delta}
\frac{1}{\pi} \frac{a}{a^2+(z_\eta-z^{\eta}_{2})^2} \to \delta(z_\eta-z^\eta_2).
\end{equation}

Finally, we comment on the role played by the term 
\begin{equation}
\mathcal{I}_n^{(b)} =- \meanOmega{\hat \psi_R(y_1,t_1)\hat \psi_R^\dagger(y_2,t_2) \hat n(x,t) },
\end{equation}
present in Eq. \eqref{eq:I} for the evaluation of the total amount of charge $\mathcal{Q}_\eta$ which travels in the direction $\eta$ after the injection. As shown in Eq. \eqref{eq:Qeta}, it is obtained integrating the chiral charge density $\delta n_\eta(z_\eta)$ over the whole system. Since one has
\begin{equation}
\int_{-\infty}^{+\infty}\!\!dx\; \hat n(x,t) \, \ket{\Omega} = 0,
\end{equation}
it turns out that all contributions to $\mathcal{Q}_\eta$ are due to $\mathcal{I}_n^{(a)}$ only. 
\section{Calculation of the total charge $\mathcal{Q}$}
\label{app:q}

In this Appendix we {calculate the total amount of charge injected in the edge channels, starting from the expression given in Eq. (\ref{eq:Qstep2}).
Let us first introduce the Fourier transform of the funcion $\beta(t)$:
\begin{equation}
\label{eq:app:beta_tilde}
\tilde \beta(E) = \int_0^{+\infty}\!\!dt \beta(t) e^{i E t} = \frac{1}{i(E-\epsilon_0)+\gamma}~.
\end{equation}
Taking advantage of the integral representation \eqref{eq:app:trasf_fourier_exp}, we write $\mathcal{Q}$ as a double integral over energies:}
\begin{equation}
\begin{split}
\mathcal{Q} &= \frac{|\lambda|^2}{2 \pi a} \frac{1}{\Gamma(A_-^2)\Gamma(A_+^2)} \left(\frac{a}{u}\right)^{1+2A_-^2} \\
&\qquad \iint_0^{+\infty} dE_1dE_2 \; E_1^{A_-^2}E_2^{A_-^2-1} \left|\tilde w\left(\frac{E_2-E_1}{u}\right)\right|^2\\
& \qquad \qquad  \qquad \left|\tilde \beta(E_1+E_2)\right|^2 e^{-a \frac{E_1+E_2}{u}}.
\label{Q}
\end{split}
\end{equation}
Moreover, since the energy level of the dot is well defined ($\gamma \ll \epsilon_0$), {the following approximation on the function $\tilde\beta(E)$ can be used:} 
\begin{equation}
\label{eq:ansatz}
\left|\tilde \beta(E)\right|^2 = \frac{1}{\gamma^2 + (E-\epsilon_0)^2} \to \frac{\pi}{\gamma}\;  \delta (E-\epsilon_0)~.
\end{equation}
Inserting this $\delta$-function in Eq. (\ref{Q}) we are left with a single integral
\begin{equation}
\label{eq:app:Q}
\begin{split}
\mathcal{Q} &= \frac{K\gamma_0}{\gamma}\;  e^{-a \frac{\epsilon_0}{u}} \frac{1}{A_-^2\Gamma^2(A_-^2)} \left(\frac{a\epsilon_0}{2u}\right)^{2A_-^2} \\
&\qquad \int_{-1}^{+1} \!\!d\chi \; (1+\chi)^{A_-^2}(1-\chi)^{A_-^2-1} \left|\tilde w\left(-\epsilon_0\chi u^{-1}\right)\right|^2\,,
\end{split}
\end{equation}
with $\gamma_0=|\lambda|^2/(2v_{\rm F})$.
{Recalling the expression of $\gamma$ found in \eqref{eq:app:gamma_general}, we conclude that $\mathcal{Q}=1$}.

\section{Calculation of $\mathcal{I}_\mathcal{H}$}
\label{app:energy}
Here we evaluate the average function 
\begin{equation}
\mathcal{I}_\mathcal{H}=  \meanOmega{\hat  \psi_R(y_1,t_1) \left[\mathcal{ \hat H}(x,t),\hat \psi_R^\dagger(y_2,t_2)\right]}
\end{equation} 
demonstrating the validity of Eq. \eqref{eq:IH_step2}, necessary in order to evaluate the energy density fluctuations $\delta \mathcal{H}(x,t)$. In particular we have to compute functions $\mathcal{M}_\eta^{(a/b)}$, introduced in Eq. \eqref{eq:IH}. 
Focusing on $\mathcal{M}_\eta^{(a)}$ we get
\begin{equation}
\begin{split}
	\mathcal{M}_\eta^{(a)} &= \meanOmega{\hat \psi_R(y_1,t_1)\, \hat \phi_\eta(z_\eta) \,\hat \psi_R^\dagger(y_2,t_2)} \\
	= &  \; - \frac{i}{2 \pi a} \meanOmega{e^{-i\sqrt{2\pi} A_{-\eta} \hat \phi_{-\eta}(z_1^{-\eta})} e^{i\sqrt{2\pi} A_{-\eta} \hat \phi_{-\eta}(z_2^{-\eta})} } \\
	&  \partial_\nu \meanOmega{e^{-i\sqrt{2\pi} A_{\eta} \hat \phi_{\eta}(z_1^{\eta})}\;  e^{i \nu \hat \phi_\eta(z_\eta)}\; e^{i\sqrt{2\pi} A_{\eta} \hat \phi_{\eta}(z_2^{\eta})} }\Big|_{\nu=0}
\end{split}
\end{equation}
where we have used Eq. \eqref{eq:boson_eq} and the identity
\begin{equation}
\hat \phi_\eta(x,t) = -i \partial_\nu e^{i \nu \hat \phi_\eta(x,t)}\big|_{\nu=0}.
\end{equation}
By means of the Baker-Hausdorff identity, one can rewrite 
\begin{equation}
\begin{split}
	\mathcal{M}_\eta^{(a)} \, (z_\eta, z_1^\pm , z_2^\pm)&= -i A_\eta \sqrt{2\pi}\;  \mathcal{G}(z_1^\pm, z_2^\pm) \\ & \qquad  \left(G\left( \eta z_\eta - \eta z_1^\eta \right) - G\left( \eta z_2^\eta -\eta z_\eta \right)\right)\;  
	\end{split}
\end{equation}
where the bosonic Green function $G$ and the fermionic correlation function $\mathcal{G}$  
have been defined in Eq. \eqref{eq:G} and in Eq. \eqref{eq:Lambda} respectively.
 It is easy to show that $\mathcal{M}_\eta^{(b)}$ has the same expression apart from a different sign in the argument of the second bosonic Green function. As a result one has
\begin{equation}
\begin{split}
\mathcal{M}_\eta^{(a)}  + \mathcal{M}_\eta^{(b)} &= -i A_\eta \sqrt{2\pi}\;  \mathcal{G}(z_1^\pm, z_2^\pm) \;  \left[2G\left( \eta z_\eta - \eta z_1^\eta \right) \right.\\ & \qquad \left. - G\left( \eta z_\eta - \eta z_2^\eta  \right) - G\left( \eta z_2^\eta - \eta z_\eta  \right)\right].
\end{split}
\end{equation}
Eq. \eqref{eq:IH_step2} can now be readily obtained simply taking the derivative of functions $G$. 

\section{Behavior of $E_\eta$}
\label{app:ploc}
In this Appendix we present details on the energy $E_\eta$ that travels in the $\eta$ direction. Such quantity, defined in Eq. \eqref{eq:E__eta} 
is expressed as in Eq. \eqref{eq:Eeta}. 

Let us start to discuss the total energy $E=E_++E_-$. In the limit $\gamma \ll \epsilon_0$ we can approximate $|\tilde \beta(\epsilon_+)|^2 \to   \delta (\epsilon_+-\epsilon_0)\;  \pi/\gamma$, see Eq. \eqref{eq:ansatz}, writing
\begin{equation}
\begin{split}
E_\eta &= \frac{K\gamma_0}{2 \gamma} \left(\frac{K\bar a}{2 \epsilon_0}\right)^{2 A_-^2} \frac{A_\eta^2}{\Gamma{(g_\eta^-)}\Gamma(g_\eta^+)} \; e^{-K \bar a} \int_{-\epsilon_0}^{+\epsilon_0} \!\! d\epsilon_- \\
& \; \quad \;  (\epsilon_0+\epsilon_-)^{g_\eta^+-1} (\epsilon_0-\epsilon_-)^{g_\eta^--1} \left|\tilde w\left( - \epsilon_-/u\right)\right|^2.
\end{split}
\end{equation}
Recalling that $g_\eta^\pm = A_\pm^2 + (1\pm \eta)/2$,  one then has
\begin{equation}
\begin{split}
E &= \epsilon_0 \frac{K\gamma_0}{\gamma} \left(\frac{K\bar a}{2}\right)^{2 A_-^2} \frac{1}{A_-^2\Gamma^2(A_-^2)} \; e^{-K \bar a} \\&
\quad \int_{-1}^{+1} \!\!d\chi \; (1+\chi)^{A_-^2}(1-\chi)^{A_-^2-1} \left|\tilde w\left(- \epsilon_0\chi u^{-1}\right)\right|^2.
\end{split}
\end{equation}
By comparing this result with the behavior of the total charge $\mathcal{Q}$ in Eq. \eqref{eq:app:Q} we can conclude that $E=\epsilon_0 \mathcal{Q}=\epsilon_0$, since $\mathcal{Q}=1$. 
The ``universal'' limit present for local injection ($\tilde{\xi}(k)\to1$) and given in Eq.~(\ref{eq:p_loc}) is recovered using in Eq.~(\ref{eq:Eeta}) the relation
\begin{equation}
\begin{split}
&\int_{-\epsilon_+}^{+\epsilon_+} \!\! d\epsilon_- \; \frac{(\epsilon_++\epsilon_-)^{g_\eta^+-1} (\epsilon_+-\epsilon_-)^{g_\eta^--1}}{\Gamma{(g_\eta^-)}\Gamma(g_\eta^+)} = \\
\qquad & = \frac{(2\epsilon_+)^{1+2A_-^2}}{\Gamma(2+2A_-^2)}~.
\end{split}
\end{equation}
We therefore  have that $E_\eta = A_\eta^2 \mathcal{E}$ with 
\begin{equation}
\begin{split}
\mathcal{E} &= \frac{K\gamma_0}{\pi} \left(\frac{K\bar a}{\epsilon_0}\right)^{2 A_-^2} \!\! \frac{1 }{\Gamma(2+2A_-^2)}\\
&\quad \int_0^{\infty}\!\!d\epsilon_+\; \epsilon_+^{1+2A_-^2} \left|\tilde \beta (\epsilon_+)\right|^2 e^{-K \bar a \frac{\epsilon_+}{\epsilon_0}}
\end{split}
\end{equation}\\
independent of $\eta$. Such an expression immediately allows to recover the energy partitioning factors $p_\eta^{loc}$ given in Eq. \eqref{eq:p_loc}.


\begin{thebibliography}{43}
\bibitem{notecutoff} Note that all physical observables are obtained in the limit $a\to 0$, where one can set $\hbar v_{{\rm F}}/a$ as the highest energy scale in the problem, of the order of the energy gap for a 2DTI ($20-30$ meV).
\bibitem{footnote:C_n} It is possible to show that, for $K=1$ and in the limit $\epsilon_0 \to \infty$ (where one can actually forget about Fermi sea) function $C_1(\tau)$ does not contribute to $j_+(\tau)$ since one has $\Re\left[C_1(\tau)\right] \sim 1$.
\bibitem{footnote:H} This can be easily shown since one has 
$\int_{-\infty}^{+\infty} \!\! dx \;  \mathcal{\hat H}(x,t) = \hat H$ and $\hat H \ket{\Omega} = 0$. As a result $\int dx \; \mathcal{I}_\mathcal{H}^{(b)} = 0$.
\end{thebibliography}
\end{document}